\documentclass[preprint,eqsecnum,aps,prd,nofootinbib]{revtex4}

\usepackage[pdftex]{color,graphicx}
\usepackage{enumerate}
\usepackage{amsmath}
\usepackage{amsfonts}
\usepackage{verbatim}

\def\be{\begin{equation}}
\def\bea{\begin{eqnarray}}
\def\ee{\end{equation}}
\def\eea{\end{eqnarray}}
\usepackage{hyperref}

\begin{document}

\title{Finite-mass correction to 2D Black-hole evaporation rate}

\author{Liora Dori and Amos Ori}

\affiliation{Department of Physics \\ Technion-Israel Institute of Technology \\ Haifa 3200, Israel }

\begin{abstract}
We numerically analyze the evolution of a two-dimensional dilatonic
black hole, within the CGHS model. We focus our attention on the finite-mass corrections to the universal evaporation rate which applies at the large-mass limit. Our numerical results confirm a previous theoretical prediction for the first-order ($\propto1/M$)
correction. In addition, our results strongly suggest that
the next-order ($\propto1/M^{2}$) term vanishes, and provide a rough
estimate for the third-order term.
\end{abstract}

\date{\today}

 \maketitle

\section{Introduction}

In the semiclassical theory of gravity, macroscopic non-spinning black
holes (BHs) emit a thermal radiation corresponding to the Hawking
temperature $T_{H}=\hbar c^3/(8\pi k_BGM)$ \cite{Hawking}. This amounts to an outflux rate $\dot{E}$ which is strictly proportional to $1/M^2$.

However, this simple and universal result is expected to hold only
at the macroscopic limit ($M\rightarrow\infty$), and one may anticipate
a finite-mass correction. The origin of this correction may be understood
as follows: To derive the quantum outflux one has to analyze the (backward)
propagation of the field's modes on the BH background, from future
null infinity (FNI) back to past null infinity (PNI). Hawking's
original derivation uses the \emph{classical Schwarzschild geometry}
as the background metric (over which the field's modes are propagated).
This is a reasonable approximation as long as the BH is very massive
(compared to the Planck mass $M_{Pl}$). However, in principle, one
should instead use the self-consistent \emph{semiclassical} geometry
as the background BH metric. The smaller the BH mass $M$, the larger
is the expected deviation of the semiclassical geometry from the classical
Schwarzschild solution. Correspondingly, one should expect a finite-mass
correction to the universal Hawking outflux, which grows with decreasing
$M$.

It may be of interest to evaluate this finite-mass correction to the
semiclassical outflux. For example, it has been argued \cite{APR}
(in the framework of two-dimensional gravity) that this correction
actually reveals the fundamental non-thermal character of the semiclassical
outflux. Such deviations from thermality could be highly relevant
to the attempts to estimate the possible amount of correlations between the
emitted particles (that is, the amount of  ``information'' encoded
in Hawking radiation). However, it is difficult to calculate such
finite-mass corrections in the realistic four-dimensional (4D) context.
The reason is that there is no known general expression for the renormalized
stress-energy tensor $\hat{T}_{\alpha\beta}$ in 4D, which makes it
hard to construct the semiclassical BH geometry (over which the quantum
field's modes are to be propagated).

The situation is much simpler in the two-dimensional (2D) context,
however. Callan, Giddings, Harvey and Strominger (CGHS) \cite{CGHS}
introduced a formalism of 2D gravity in which the metric is coupled
to a dilaton field $\phi$ and to a large number $N$ of identical
massless scalar fields. In this 2D framework $\hat{T}_{\alpha\beta}$
is known explicitly, allowing one to translate semiclassical dynamics
into a closed system of partial differential equations (PDEs) \cite{CGHS}.
Although the exact solution to these PDEs is not known, certain approximate
solutions have been derived \cite{Aprox}. Also, it is possible to
numerically integrate these PDEs and thereby to explore 2D semiclassical
dynamics \cite{APR,previous}. Among other things, such a numerical
integration allows one to analyze the rate of evaporation for finite
BH mass as well.

There is a remarkable difference between 2D and 4D classical BHs:
Whereas in the latter the horizon's surface gravity $\kappa$ scales
as $1/M$, in 2D it is \emph{independent of the BH mass}. As a consequence,
the (large-$M$) Hawking temperature is constant (i.e. independent
of $M$) in the 2D framework, and so is the outflux $\dot{E}$ \cite{CGHS}.
This contrasts with the 4D case, wherein $\dot{E}\propto M^{-2}$.
Though, for the same reasons explained above, this 2D constant outflux
only holds in the macroscopic limit, and one should anticipate finite-mass
corrections.

In Ref. \cite{Aprox} an approximate solution to the CGHS field equations
was constructed, accurate to first order in $N/M$. Based on this
approximate geometry, it is possible to derive \cite{Aprox1} the
leading-order ($\propto M^{-1}$) finite-mass correction to the 2D
constant outflux. The corrected value was found to be
$\dot{E}=K[1/4+c_{1}K/M_B+O(M_B^{-2})]$
(in appropriate units; see below). Here $K\equiv N/12$, $c_{1}>0$
is a certain known coefficient \cite{Aprox1}, and $M_B$ denotes
the Bondi mass, which is essentially the remaining BH mass (see section
3 for more details). Note that the outflux increases with time, because
$M_B$ steadily decreases.

One of the main goals of this paper is to numerically explore this
finite-mass correction to $\dot{E}$, and to verify the aforementioned
$O(1/M_B)$ theoretical prediction.

An independent numerical integration of the CGHS system has been carried
out recently by Ashtekar, Pretorius and Ramazanoglu (APR) \cite{APR}.
They also explored numerically the dependence of the outflux on the
BH mass. However, their investigation was restricted to the range
of relatively small masses, where the relevant $\propto1/M_B$ expansion
parameter is not quite $\gg1$ (which makes it harder to interpret
$\dot{E}$ in terms of inverse powers of $M_B$). In the numerical
analysis presented here we significantly increase the BH mass, by
a factor $2.5$. This allows us to carry out a more detailed analysis
of $\dot{E}$ in terms of inverse powers of $M_B$.

When we numerically obtained the (time-dependent) value of $\dot{E}$,
we were striked by its remarkable similarity to the above-mentioned
first-order corrected theoretical prediction. In fact, it was not
possible to visually distinguish between the numerical and theoretical
curves (see Fig. 1 below). This came to us as a surprise, because
one may naturally expect to have $O(1/M_B^{2})$ corrections as
well --- and no such corrections can be seen in Fig. 1. This could
hint that the coefficient of the $O(1/M_B^{2})$ term in the outflux
function $\dot{E}(M_B)$ (which has not yet been derived analytically)
may actually vanish. This unexpected observation motivated us to analyze
$\dot{E}(M_B)$ in more detail, in order to get a better insight
into the $O(1/M_B^{2})$ term (and possibly also into the $O(1/M_B^{3})$
term).

Our detailed numerical results strongly support the conjecture that
the $O(1/M_B^{2})$ term indeed vanishes, and also provide a crude
estimate of the (non-vanishing) $O(1/M_B^{3})$ term.

\section{The model and field equations}

The CGHS model \cite{CGHS} consists of a two-dimensional metric $g_{\alpha\beta}$
coupled to a dilaton $\phi$ and to a large number $N\gg1$ of identical
massless scalar fields $f_{i}$. We express the metric in the double-null
form, namely $ds^{2}=-e^{2\rho}du dv$. The action then reads
\begin{equation}
\label{Action}
\frac{1}{\pi}\int{dudv \left[{{e^{-2\phi}}(-2{\rho_{,uv}}+4{\phi_{,u}}{\phi_{,v}}-{\lambda^2}{e^{2\rho}})-
\frac{1}{2}\sum\limits_{i=1}^N{{f_{i,u}}{f_{i,v}}}+\frac{N}{{12}}\rho,_{u}\rho,_{v}}\right]}.
\end{equation}
The last term in the action expresses the semiclassical effects, derived from the trace anomaly.

The model also contains a cosmological constant $\lambda^{2}$. Throughout
this paper we set $\lambda=1$. This choice (along with $\textcolor[rgb]{0.00,0.00,1.00}{c}=G=1$)
fully determines the system of units, making all variables dimensionless
\footnote{ Formally it is equivalent to the change of variable
$\rho'=\rho+\ln(\lambda)$,
which does not affect the field equations otherwise. Note that hereafter
we also set $f_{i}=0$ in the field equations, as we are dealing here
with the evaporation of the BH rather than its formation.}.

To simplify the field equations we introduce new variables (following
Refs. \cite{2Dcharged,Aprox}): $R\equiv e^{-2\phi}$ and $S\equiv2(\rho-\phi)$.
In these variables the model's evolution equations take the form
\begin{eqnarray}
 & R,_{uv}=-e^{S}-K\rho,_{uv},\label{evolution}\\
 & S,_{uv}=K\rho,_{uv}/R,\nonumber
 \end{eqnarray}
where $K\equiv N/12$  [and $\rho=(S-\ln R)/2$ is to be substituted ].
There are also two constraint equations:
\begin{eqnarray}
R,_{ww}-R,_{w}S,_{w}+\hat{T}_{ww}=0,\label{constraint}
\end{eqnarray}
where hereafter $w$ stands for either $u$ or $v$, and $\hat{T}_{ww}$
is the ($ww$) component of the renormalized stress-energy tensor
$\hat{T}_{\alpha\beta}$. From the trace anomaly one obtains \cite{CGHS}
(via energy-momentum conservation) an explicit expression for $\hat{T}_{ww}$:
\begin{eqnarray}
\hat{T}_{ww}=K\left[{\rho,_{ww}-\rho^{2},_{w}+z_{w}\left(w\right)}\right],\label{eq:fluxes}
\end{eqnarray}
where $z_{w}\left(w\right)$ is a certain boundary function (to be
determined from the initial conditions).

Setting $K=0$, one recovers the classical evolution equations
\begin{equation}
R,_{uv}=-e^{S},\quad S,_{uv}=0,
\label{ClassEquations}
\end{equation}
 and the constraint equations $R,_{ww}=R,_{w}S,_{w}$. This set of
equations admits a one-parameter family of classical solutions  [up
to gauge transformations of the general form $u\rightarrow u'(u)$,
$v\rightarrow v'(v)$ ], which is the two-dimensional analog of the
standard Schwarzschild solution. In the so-called \emph{Eddington
coordinates} this solution takes the simple form
\begin{equation}
R=M+e^{v-u},\quad S=v-u,
\label{SchwartzEddington}
\end{equation}
where $M$ is a constant. For each $M>0$ the solution describes a
static BH with mass $M$. Note that in the classical solution
\begin{equation}
\rho(u,v)=-\frac{1}{2}\ln(1+Me^{u-v})\quad\quad\mbox{(classical).}
\label{eq:Classical}
\end{equation}
This implies asymptotic flatness ($\rho\rightarrow0$) at both PNI
($u\rightarrow-\infty$) and FNI ($v\rightarrow\infty$). The special
case $M=0$ (also known as the \emph{linear dilaton} solution) describes
a flat spacetime, $\rho(u,v)=0$.

Next we consider the collapse of a thin shell of mass $M>0$. Following
CGHS, we assume that the shell propagates along an ingoing null line,
which we set to be $v=0$. At the classical level the solution is
(\ref{SchwartzEddington}) at $v>0$ and flat at $v<0$.
\footnote{This flat solution takes the form $R=e^{v}(M+e^{-u}),$ $S=v-u$.
 [A gauge transformation $u\to u'=-\ln(M+e^{-u})$ will then bring it to its more standard form $R=e^{v-u'}, S=v-u'$.]}
At the semiclassical level the geometry is still flat at $v<0$.
However at $v>0$ the classical geometry (\ref{eq:Classical}) is now
replaced by a corresponding solution of the semiclassical field equations
(\ref{evolution}, \ref{constraint}). The characteristic initial data
for the semiclassical evolution equations may conveniently be prescribed
on the collapsing shell and along PNI: These are exactly the same
initial data as in the classical collapsing-shell solution \cite{Initial}.

\section{Outflux at FNI}

Throughout the rest of the paper we shall use $u$ and $v$ to denote
the Eddington-like coordinates for the semiclassical solution at $v>0$.
They are defined by the requirement that the solution takes the asymptotic
form (\ref{SchwartzEddington}) at both (left) PNI and FNI (and in addition
the collapsing shell is located at $v=0$).

The shell collapse triggers the onset of \emph{Hawking radiation},
namely a nonvanishing energy outflux $T(u)\equiv T_{uu}(u,v\rightarrow\infty)$
at FNI (it is the same quantity that we denoted by $\dot{E}$ throughout
the Introduction)
\footnote{Note that the outflux $T$ is specifically defined here as the $(uu)$
component of $\hat{T}_{\alpha\beta}$ with respect to the \emph{ Eddington
coordinate} $u$ (i.e. the one for which $ $$2g_{uv}\rightarrow-1$
at FNI).
}. It may be obtained from Eq. (\ref{eq:fluxes}) by \begin{equation}
T(u)=\hat{T}_{uu}(u,v\rightarrow\infty)-\hat{T}_{uu}(u,v=0)\label{eq:general}\end{equation}
 [a convenient combination which cancels out the boundary function
$z_{u}\left(u\right)$ ], because presumably no outflux crosses the
collapsing shell. This expression depends on the actual solution $\rho(u,v)$
at $v\geq0$ through the RHS of Eq. (\ref{eq:fluxes}).

In the macroscopic limit ($M\rightarrow\infty$), one may substitute
the classical geometry (\ref{eq:Classical}) for $\rho(u,v)$, obtaining
the leading-order expression \begin{equation}
T(u)\approx\frac{K}{4}\left[1-\frac{1}{(1+Me^{u})^{2}}\right]\equiv T_{(0)}^{gl}\quad\quad\mbox{(zero-order; global).}\label{eq:ZeroFull}\end{equation}
We shall denote the term in squared brackets by $F(u)$ for brevity,
and refer to it as the \emph{transition function}, describing the
onset of Hawking radiation. It starts from zero at early $u$, and
quickly approaches the asymptotic value $1$ at large positive $u$. Thus, the
outflux quickly approaches the constant asymptotic value
\begin{equation}
T\approx\frac{K}{4}\equiv T_{(0)}^{late}\quad\quad\mbox{(zero-order; asymptotic)}
\label{eq:ZeroLate}
\end{equation}

This zero-order calculation properly describes the outflux at the
large-$M$ limit. However, for a finite-mass BH the expression for
$T$ is modified because the semiclassical $\rho(u,v)$ differs from
the classical solution (\ref{eq:Classical}). The leading-order semiclassical
correction to the classical solution for $R$ and $S$ (and hence
for $\rho$) has been analyzed in Ref. \cite{Aprox}. Based on this,
the leading-order correction to the outflux was calculated in \cite{Aprox1}.
The asymptotic (i.e. late-$u$) result was found to be
\begin{equation}
T(u)\simeq\frac{K}{4}\left[1+\frac{K}{2M_B(u)}\right]\equiv T_{(1)}^{late}\quad\quad\mbox{(1st-order; asymptotic)}\label{eq:1stLate}\end{equation}
where\begin{equation}
M_B(u)=M-\intop_{-\infty}^{u}T(u')du'\label{eq:Mbondi}\end{equation}
is the Bondi mass. (Essentially $M_B(u)$ denotes the remaining
BH mass as ``seen'' by a distant observer \cite{Bondi_mass}.)

Finally, by combining the asymptotic result (\ref{eq:1stLate}) with
the transition function $F(u)$, we arrive at the global, first-order
corrected, expression for the outflux: \begin{equation}
T(u)\simeq\left[1+\frac{K}{2M_B(u)}\right]T_{(0)}^{gl}\equiv T_{(1)}^{gl}\quad\quad\mbox{(1st-order; global)}\label{eq:1stFull}\end{equation}
We shall shortly verify this approximate expression for $T(u)$
by comparing it to a numerical simulation.

The global expressions $T_{(0)}^{gl}$ and $T_{(1)}^{gl}$ are obviously
more effective than their respective late-time counterparts $T_{(0)}^{late}$
and $T_{(1)}^{late}$, as they properly describe the transient stage
of the onset of Hawking radiation. It should be pointed out, though,
that the simpler, late-time asymptotic expressions $T_{(0)}^{late}$
and $T_{(1)}^{late}$ also have their advantage: They serve as ``universal curves'' \cite{APR} for $T(M_B)$ (at zeroth and 1st-order in
$K/M_B$ , respectively), onto which all evolutions with sufficiently
large initial mass should converge, regardless of initial conditions.

\section{Numerical results}

We numerically explored the semiclassical 2D spacetime of shell collapse,
using a second-order finite-difference code. The initial mass was
taken to be $M=20K$.
\footnote{The semiclassical CGHS model admits an exact scaling law in which
$K$ changes (that is, the number of scalar fields changes), and at
the same time the various model's variables are rescaled by certain
powers of $K$ \cite{Aprox,APR}. In particular, the shell's mass
$M$, the Bondi mass $M_B$, and the outflux $T$ all scale as $K$.
The functional dependence of $T/K$ on $M_B/K$ is thus invariant
to this rescaling. This scaling law allows one to obtain results for
all $K$ values from numerical integrations with a single fiducial
$K$ value, e.g. $K=1$. }
(This is to be compared with the initial value $M=8K$ used by APR
\cite{APR}, and much smaller values $M\lesssim K$ 
used in earlier analyses \cite{previous}). Full details
of the analysis will be presented elsewhere \cite{Dori}. The domain
of integration covers the range of $u$ wherein the Bondi mass decreases
from its original value $M_B=20K$ up to $M_B\approx15K$ (at
larger $u$ --- i.e. smaller $M_B$ --- numerical errors start to
grow exponentially). We calculated $T(u)$ by Eq. (\ref{eq:general}),
\footnote{In practice we have calculated $\hat{T}_{uu}(u)$ along several lines
of constant $v$, at sufficiently large $v$ values which mimic FNI.
These different $v=const$ lines are indistinguishable in the figures below (see
also footnote \ref{numerics}). }
and then constructed $M_B(u)$ from it, through Eq. (\ref{eq:Mbondi}).

Figure 1 displays the numerically-obtained function $T(u)$, compared
with the global first-order theoretical prediction $T_{(1)}^{gl}(u)$. Both functions $T$ and $T_{(1)}^{gl}$ are plotted against the Bondi mass $M_B(u)$.
For reference, the simpler (but less precise) approximate expressions
$T_{(0)}^{late}$, $T_{(0)}^{gl}$, and $T_{(1)}^{late}$ are also
shown. Remarkably, the numerical curve is visually \emph{indistinguishable}
from $T_{(1)}^{gl}(u)$, even in the zoomed figure 1b. This observation
came to us as a surprise, because it would be just natural to expect
corrections to $T_{(1)}^{gl}(M_B)$ of order $O(K/M_B)^{2}$.
The two grey curves in Fig.(1b) represent our original expectation for the
typical order of magnitude of such a putative second-order correction
term.
\footnote{The coefficient of this second-order term has not been analytically calculated so far. The gray curves in Fig.\ 1a represent a fiducial value, obtained by naively extending the (known) zeroth and first-order coefficients to the next order as a geometric progression. }
The graph shows no signature of such a correction term. In particular
Fig. 1b indicates that if such a second-order term at all exists,
it must be $\ll$ than its naively-expected order of magnitude.

\begin{figure}[ht]
\begin{center}
\includegraphics[scale=.35]{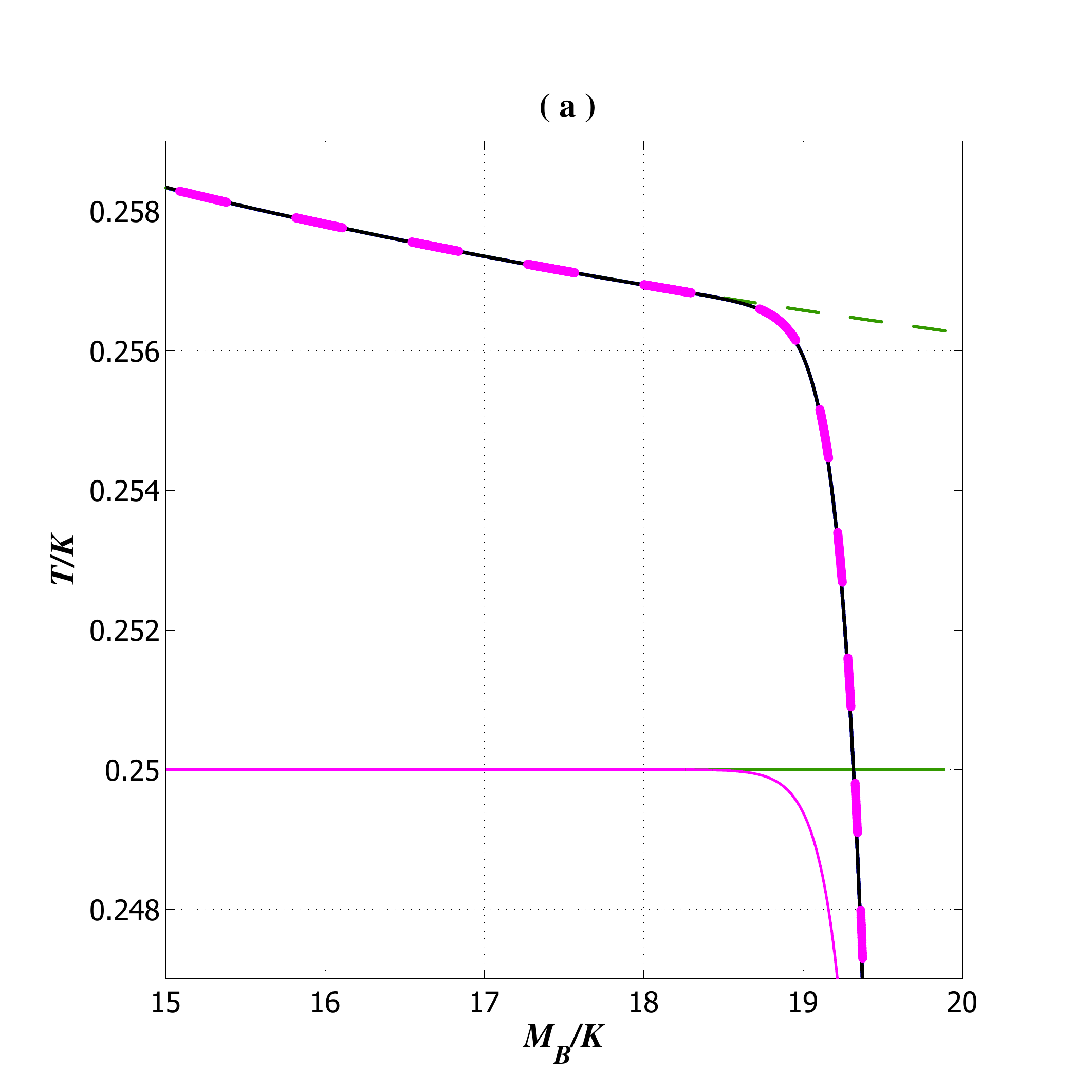}
\includegraphics[scale=.35]{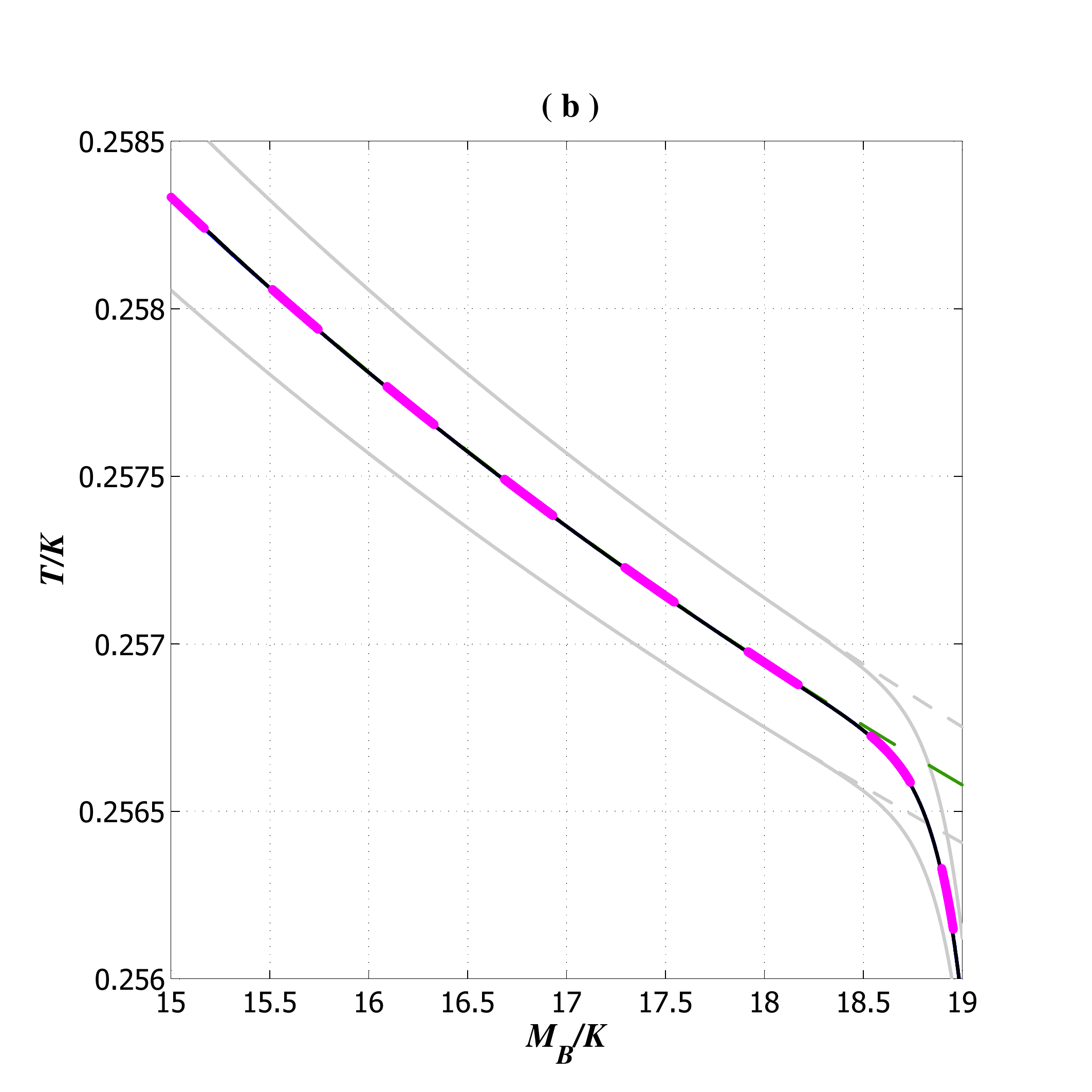}
\caption{ \label{summary}
Comparison of our numerical results for the outflux $T(u)$ (solid
black curve), as a function of $M_B(u)$, to certain approximate analytical expressions. 
(a): The two pink curves represent the (global) zeroth-order and first-order theoretical predictions, $T_{(0)}^{gl}$ (solid) and $T_{(1)}^{gl}$ (dashed). 
The two green curves represent the corresponding late-time asymptotic expressions $T_{(0)}^{late}$ (solid) and $T_{(1)}^{late}$ (dashed). 
(b): A blow-up on Fig. 1a. 
The additional light-gray curves represent one's naive expectation for the typical magnitude of a putative second-order correction term (see main text and footnote therein). 
Note that the numerical curve is visually indistinguishable from
$T_{(1)}^{gl}(u)$, even in the zoom level of Fig. 1b.
(The solid black curve actually displays numerical data taken along four different $v=const$ lines throughout the range $24\le v\le 30$, which are again visually indistinguishable.)}
\end{center}
\end{figure}

This observation led us to suspect that perhaps there actually is
no second-order finite-mass correction to $T_{(1)}^{gl}$ and $T_{(1)}^{late}$.
Furthermore, it provoked the intriguing possibility that perhaps the
first-order corrected expression $T_{(1)}^{late}$ is the \emph{exact}
expression for the outflux (in the asymptotic late-time limit).

To address these questions, we explored the residual
$\Delta T\equiv T-T_{(1)}^{gl}$
with a much higher zoom level. Figure 2 displays $\Delta T$ as a function
of $M_B$  \footnote{For the inspected mass-range, $17.3\leq M_B \leq 18$, $T_{(1)}^{gl}$ has already approached its asymptotic form $T_{(1)}^{late}$. Therefore we can use the residual $\Delta T$ defined above to study the next-order correction term for $T^{late}$.}. Note the tiny vertical scale $\sim10^{-6}$. At such
a small scale, the truncation error becomes a significant issue. The
numerical simulation used a grid step-size of $0.0025$ (in both Eddington coordinates $u$ and $v$). We also carried out simulations with larger step sizes $0.005$ and $0.01$, and verified second-order
convergence (in a certain range of $M_B$, displayed in Fig. 2).
We then used Richardson extrapolation to correct the truncation error
in our finest run (step-size $0.0025$). It is the corrected residual
$\Delta T$ which is shown in Fig. 2.
\footnote{
\label{numerics}
Beside the finite step-size (and round-off) there are two other potential
sources of errors in our simulations: (i) We start the simulation
at a finite $u=u_{0}$ (rather than at PNI, $u_{0}\rightarrow-\infty$);
(ii) we compute $\hat{T}_{uu}$ at finite $v=v_{final}$ (rather than
at FNI, $v_{final}\rightarrow\infty$). The finite-$u_{0}$ error
is corrected by the combination of two methods: First, we apply first-order
weak-field semiclassical correction to the (otherwise classical) initial
data that we set at $u=u_{0}$ \cite{Dori}; Second, we run the code
with three different $u_{0}$ values and perform a Richardson extrapolation.
The finite-$v_{final}$ error is handled by trying several different
values of $24\leq v_{final}$$\leq 32$ and verifying that our results
for $\Delta T$ are unaffected by further increasing $v_{final}$. }
\begin{figure}[ht]
\begin{center}
\includegraphics[scale=.50]{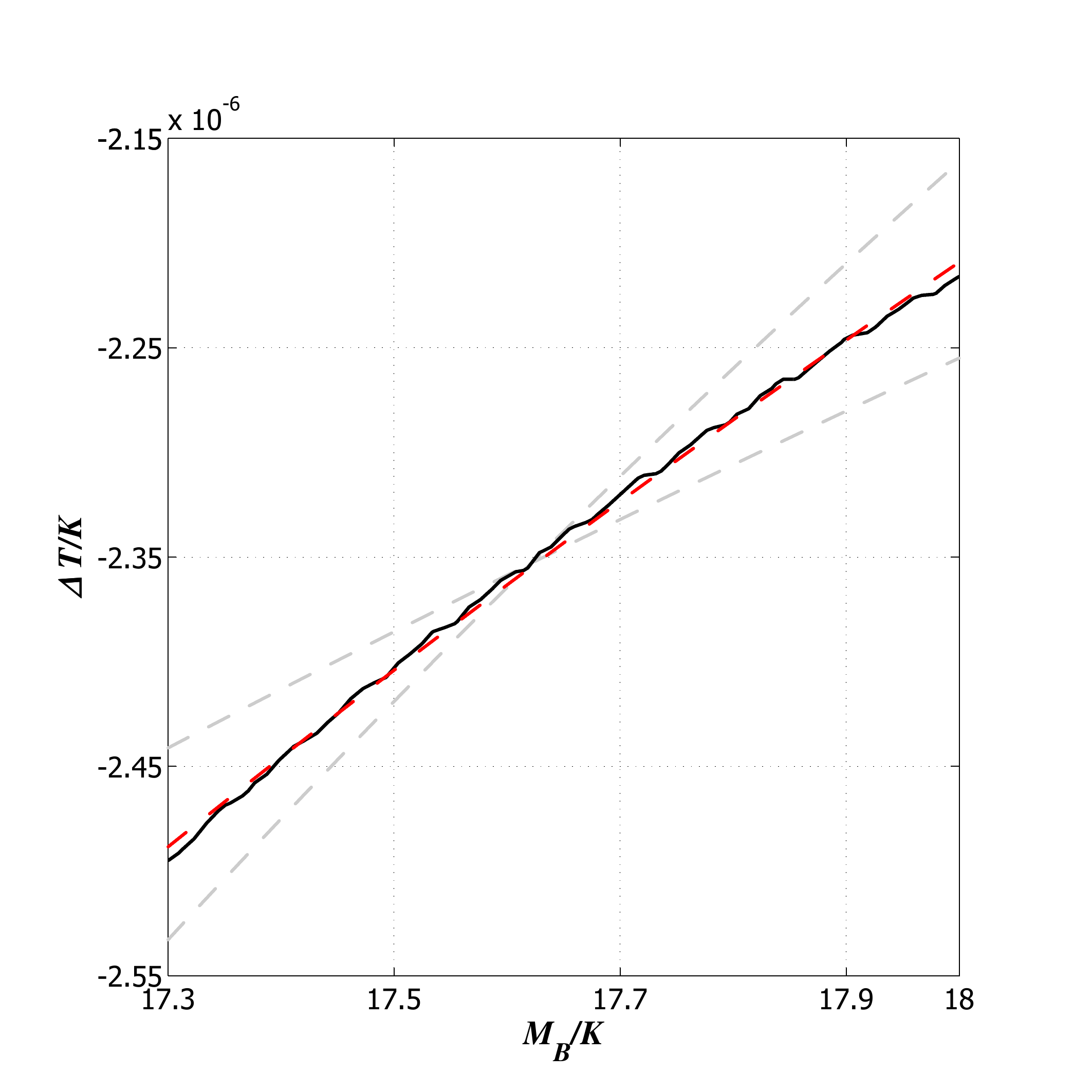}
\caption{ \label{fitting}
The residual $\Delta T$ (solid black curve) as a function of the Bondi mass $M_B$. The matched curve $\Delta T_{(3)}$ ($\propto M_B^{-3}$) is shown by dashed red line. 
For comparison we also display two additional curves (dashed gray) proportional to $M_B^{-2}$ and $M_B^{-4}$. 
(The wiggles in the solid numerical curve result from the round-off error.)}
\end{center}
\end{figure}

Two observations emerge from this figure: First of all, a residual
$\Delta T$ certainly exists  [that is, the first-order corrected
expression $T_{(1)}^{late}(M_B)$ is \emph{not} the exact (late-time,
asymptotic) expression for $T$ --- an issue which was left open in
Fig. 1b ]. Nevertheless, Fig. 2 is also very suggestive that a second-order
correction probably does not exist. The residual $\Delta T$ is well
matched by a term $-0.05(K/4)(K/M_B)^{3}\equiv \Delta T_{(3)}$, though with a relatively
large numerical uncertainty in the pre-factor, which we estimate as
$\sim\pm25\%$. (This large relative uncertainty is obviously attributed to the tiny overall magnitude of the residual, which is smaller than
$T$ by a factor $\sim10^{-5}$. 
\footnote{The main source of this uncertainty is the finite $u_0$ (see footnote \ref{numerics}). Since we only used three $u_0$ values, it is hard to assess the effectiveness of the associated Richardson extrapolation.})

\section{Summary}

Our numerical results confirm the previous theoretical prediction
of the first-order finite-mass correction (\ref{eq:1stLate},\ref{eq:1stFull}).
They further suggest the absence of a second-order correction term,
and provide a rough estimate for the third-order term. Our final
result is the following third-order approximate expression for $T(M_B)$:
\begin{equation}
T_{(3)}^{late}=\frac{K}{4}\left[1+\frac{K}{2M_B}+c_{3}\left(\frac{K}{2M_B}\right)^{3}\right]\label{eq:3rdLate}\end{equation}
along with its global counterpart $T_{(3)}^{gl}=F(u)\, T_{(3)}^{late}$.
Here $c_{3}$ is a dimensionless coefficient which we estimate as
$\sim-0.4$ (with about $\sim 25\%$ uncertainty).

It would be desired to find the analogous finite-mass corrections
to the Hawking outflux from a 4D semiclassical BH, but this is obviously
a much harder task.

\section*{Acknowledgment}

We would like to thank Abhay Ashtekar, Frans Pretorius, and Fethi
Ramazanoglu for helpful discussions. This research was supported by
the Israel Science Foundation (grant no. 1346/07)

{\it Note added in Proof:}Ê
After this work was submitted, Ramazanoglu notified us that he managed to convert the results of Ref. \cite{APR} for the mass-dependent outflux, from the modified definitions of ÓoutfluxÓ and ÓBondi massÓ (introduced in Ref. \cite{ATV}) to the traditional ones \cite{Bondi_mass}.Ê HeÊ then carried out an expansion of the (traditional) outflux in inverse powers ofÊ
(traditional) Bondi mass. His results, truncated after third order, agree with our 
Eq. (\ref{eq:3rdLate}) with $c_3=-(5/12) \approx -0.42$  \cite{Fethi}. 



\begin{thebibliography}{11}
\bibitem{Initial} This set-up of semiclassical initial data is described
in detail in Ref. \cite{Aprox}.
\bibitem{Bondi_mass} This is the traditional definition of the
Bondi mass. In Ref. \cite{ATV} new definitions were introduced for
both the ``outflux'' and the ``Bondi mass''. These new definitions
admit some elegant properties. Here we prefer to use the traditional
definitions for the outflux and $M_B$, because these traditional
definitions make use only of the local geometric (and dilatonic)
behavior at FNI (whereas the definitions introduced in \cite{ATV}
also resort to the time-translation coordinate at left-PNI).
\end{thebibliography}
\end{document}